\documentclass[prl,showpacs,showkeys,nofootinbib,twocolumn,floatfix,superscriptaddress]{revtex4}

\usepackage{graphicx}  %
\usepackage{amsmath}
\usepackage{bm}  %
\usepackage{ulem}

\newcommand{\bea}{\begin{eqnarray}}
\newcommand{\eea}{\end{eqnarray}}
\newcommand{\beq}{\begin{equation}}
\newcommand{\eeq}{\end{equation}}
\newcommand{\bqa}{\begin{eqnarray}}
\newcommand{\eqa}{\end{eqnarray}}

\def\mqo2{{\!\!\!}}

\begin{document}

\title{Short-Time Operator Product Expansion\\
 for rf Spectroscopy of a Strongly-interacting Fermi Gas}

\author{Eric Braaten}
\affiliation{Department of Physics,
         The Ohio State University, Columbus, OH\ 43210, USA\\}

\author{Daekyoung Kang}
\affiliation{Department of Physics,
         The Ohio State University, Columbus, OH\ 43210, USA\\}
\author{Lucas Platter}
\affiliation{Institute for Nuclear Theory,
University of Washington, Seattle WA\ 98195 USA\\}
\date{\today}

\begin{abstract}
Universal relations that hold for any state 
provide powerful constraints on systems consisting of fermions 
with two spin states interacting with a large scattering length.
In radio-frequency (rf) spectroscopy, the mean shift in the rf frequency
and the large frequency tail of the rf transition rate are proportional 
to the contact, which measures the density of pairs with small separations.
We show that these universal relations can be derived 
and extended by using the short-time operator product expansion 
of quantum field theory.  This is a general method for identifying 
aspects of many-body physics that are controlled by few-body physics.
\end{abstract}

\smallskip
\maketitle

Trapped ultracold atoms allow the study of the few-body physics 
and the many-body physics of systems in which the fundamental 
interactions between the constituents are understood 
and can be controlled experimentally.
The possibility of making the interactions between the atoms 
arbitrarily strong presents a challenge to theory, 
because many theoretical methods break down when interactions 
become too strong.  
Specifying the interactions between the constituents 
of the many-body system in terms of 2-body interactions
is a trivial example of a connection between few-body physics and
many-body physics, but there can be deeper connections.
Among the simplest examples of strongly-interacting systems are ones 
that consist of atoms that interact 
with a large scattering length \cite{Braaten:2004rn}. 
The 2-body problem for such atoms can be solved analytically, 
the 3-body problem can be solved exactly numerically,
and the 4-body problem is becoming tractable using modern computers.
This makes it an interesting system for studying how nontrivial aspects of 
many-body physics can be controlled by few-body physics.

A powerful experimental tool for studying ultracold atoms is 
radio frequency (rf) spectroscopy, in which atoms in one 
hyperfine spin state are excited into a different spin state.
Pioneering applications of this method were measurements 
of the binding energy of weakly-bound diatomic molecules of 
$^{40}$K atoms \cite{jin0305}
and the study of the pairing gap in a many-body system of 
$^6$Li atoms \cite{grimm0405}.
The method had been extended to allow the spacial resolution 
of trapped systems \cite{ketterle0705}
and the momentum resolution of the excited atoms \cite{jin0805}.
A recent review of rf spectroscopy has explored the relation 
to photoemission experiments on high-temperature superconductors
\cite{levin0810}.

Universal relations between various properties of a system 
that must be satisfied in any state can provide powerful 
constraints on theoretical methods.
Shina Tan has derived universal relations for systems consisting 
of fermions with two spin states that interact with a 
large scattering length $a$ \cite{Tan}.
These relations all involve a property of the system 
called the {\it contact}.
It is an extensive quantity that can be expressed as the integral 
over space of the {\it contact density},
which is proportional to the number of pairs with different spins 
per (volume)$^{4/3}$ \cite{Braaten:2008uh}.   
The Tan relations include the coefficient of the $1/k^4$ tail
in the momentum distributions at large momentum, 
a decomposition of the total energy into terms that are insensitive 
to short distances, the rate of change of the 
free energy due to a change in $a$,
the relation between the pressure and the energy density 
in a homogeneous system, and the virial theorem 
for a system in a harmonic trapping potential \cite{Tan}.
Tan derived his relations within the framework of the many-body 
Schr\"odinger equation using novel methods involving generalized functions.
In Ref.~\cite{Braaten:2008uh}, Braaten and Platter rederived the Tan relations
within the framework of quantum field theory using standard  
renormalization methods together with the 
{\it short-distance operator product expansion}.
The Tan relations have also been rederived using 
less formal methods \cite{other}.

The contact also plays an important role in the radio-frequency (rf) 
spectroscopy of ultracold atoms.  In a many-body system 
consisting of atoms in spin states 1 and 2, an rf signal 
with frequency $\omega$ can be used to transfer atoms
from the spin state 2 into a third spin state 3.
There are sum rules that constrain the rf transition rate 
$\Gamma(\omega)$ \cite{YB05}.
Of particular interest is the case of fermionic atoms for which every pair
has strong interactions 
determined by large scattering lengths $a_{12}$, $a_{13}$, and $a_{23}$.
An example is the lowest three hyperfine spin states of $^6$Li atoms 
at generic values of the magnetic field.
In this case, the sum rules are
\begin{subequations}
\begin{eqnarray}
\int_{-\infty}^\infty \!\!\! d \omega~\Gamma(\omega)
&=& \pi \Omega^2 N_2,
\label{sumrule-0}
\\
\int_{-\infty}^\infty \!\!\! d \omega ~\omega~\Gamma(\omega)
&=& (\Omega^2/4m) (a_{12}^{-1} - a_{13}^{-1}) C_{12},
\label{sumrule-1}
\end{eqnarray}
\label{sumrules}
\end{subequations}
where $\Omega$ is the Rabi frequency determined 
by the strength of the rf signal,
$N_2$ is the number of atoms of type 2,
and $C_{12}$ is the contact for atoms of types 1 and 2.
We measure the rf frequency $\omega$ relative to the hyperfine 
frequency difference between atoms 2 and 3, and we set $\hbar = 1$. 
The sum rule in Eq.~(\ref{sumrule-1}) \cite{PZ0707,BPYZ0707}
implies that the mean shift in the rf frequency 
due to interactions 
is proportional to the contact and vanishes if $a_{13} = a_{12}$.
The coefficient of the contact diverges in the limit $a_{13} \to 0$,
which indicates that the mean shift in the rf frequency 
is sensitive to the range $r_0$
of the interactions between the atoms if $a_{13}$ is not large.
The sensitivity to the range arises in this case because
$\Gamma(\omega)$ has a high-frequency tail that decreases 
like $\omega^{-3/2}$ 
and is proportional to the contact \cite{SSR0903}:
\begin{equation}
\Gamma(\omega) \longrightarrow
\frac{\Omega^2}{4 \pi \sqrt{m} \omega^{3/2}} C_{12} \qquad
(a_{13}\approx 0)~.
\label{tail-omega}
\end{equation}
The power-law tail extends out to 
$\omega$ of order $1/(m r_0^2)$,
beyond which it is cut off by range effects.

In this paper, we point out that the 
{\it short-time operator product expansion} 
of quantum field theory provides a general method for deriving 
connections between few-body physics and many-body physics.  
We use it to rederive and extend the universal results for
rf spectroscopy in Eqs.~(\ref{sumrule-1}) and (\ref{tail-omega}).
We also use it to derive new sum rules 
that are insensitive to range effects.

The {\it operator product expansion} (OPE) is a powerful tool 
for studying strongly-interacting quantum field theories that
was invented independently by Wilson, Kadanoff, and Polyakov 
in 1969 \cite{OPE}.  It expresses the product of 
local operators at different space-time points
as an expansion in local operators with coefficients 
that are functions of the separation of the operators:
\begin{eqnarray}
{\cal O}_A(\bm{R} + \mbox{$\frac12$} \bm{r},T + \mbox{$\frac12$} t)~
{\cal O}_B(\bm{R} - \mbox{$\frac12$} \bm{r},T - \mbox{$\frac12$} t) 
\nonumber \\
= \sum_C W_{C}(\bm{r},t) {\cal O}_C (\bm{R},T) .
\label{OPE-gen}
\end{eqnarray}
The sum is over the infinitely many local operators ${\cal O}_C$.  
The functions $W_C(\bm{r},t)$ are called {\it Wilson coefficients}.
The {\it short-distance} OPE, 
with the operators at the same time ($t=0$), is 
an asymptotic expansion for small $|\bm{r}|$ \cite{WZ:1972}.
It can be generalized to a {\it short-time} OPE
by analytically continuing the time difference $t$ to a 
Euclidean time $\tau$ defined by $t = - i \tau$.
This form of the OPE is an asymptotic expansion for small $|\bm{r}|$
and $\tau$.

A classic application of the short-time OPE
in high energy physics is electron-positron ($e^+ e^-$) annihilation.
In high-energy $e^+ e^-$ collisions,
most of the annihilation cross section is into hadrons (mesons and baryons).
The fundamental theory for hadrons is quantum chromodynamics
(QCD), which is a strongly-interacting quantum field theory.  
As far as the hadrons are concerned, the annihilation of an $e^+ e^-$ 
pair with center-of-mass energy $E$ results in the electromagnetic 
current operator ${\bf J}(\bm{r},t)$ acting on the QCD vacuum state.
This creates a quark and antiquark at a point with total energy 
$E$ and with equal and opposite momenta.  
They are subsequently transformed by the strong interactions of QCD 
into one or more hadrons.  The inclusive cross section $\sigma(E)$
for $e^+ e^-$ annihilation into hadrons can be expressed formally 
in terms of the expectation value in the QCD vacuum of a product of 
electromagnetic currents:
\begin{eqnarray}
\sigma(E) = (4 \pi \alpha/3 E^4)~{\rm Im} \, i \!\!
\int \!\! dt~e^{i (E + i \epsilon) t} \int \!\! d^3r 
\nonumber\\
\times
\langle 0 | {\rm T} \bm{J}(\bm{r},t) \cdot \bm{J}(\bm{0},0) | 0 \rangle,
\label{sigma-JJ}
\end{eqnarray}
where $\alpha \approx 1/137$ is the fine structure constant of QED.
The short-time OPE can be used to expand
the operator product $\bm{J}(\bm{r},t) \cdot \bm{J}(\bm{0},0)$ 
in powers of local operators with increasing dimensions.  
The leading operator is the unit operator, whose expectation value 
in any normalized state is 1 and whose dimension is 0.  
The next higher dimension operators that have nonzero expectation values 
in the QCD vacuum are scalar quark operators with dimension 3
and the gluon field strength operator with dimension 4.
Because of the asymptotic freedom of QCD, the Wilson coefficients
can be calculated using perturbation theory in the 
QCD coupling constant $\alpha_s$.  
For large complex $E$, the Fourier transforms of the Wilson
coefficients of the higher dimension operators are suppressed 
by powers of $E$.
The cross section at asymptotically large energy $E$ can therefore 
be obtained by 
truncating the OPE after the unit operator and
calculating its Wilson coefficient to leading order in $\alpha_s$.
Inserting the OPE into Eq.~(\ref{sigma-JJ}) and neglecting 
quark masses relative to $E$, one obtains the simple result
\begin{equation}
\sigma(E) \longrightarrow
4 \pi \alpha^2 \sum_q e_q^2/E^2,
\label{sigma-OPE}
\end{equation}
where the sum is over the quark flavors $q$ and 
$e_q$ is the electric charge ($+\frac23$ or $-\frac13$) of the quark
\cite{Appelquist:1973uz}.
The leading corrections at large $E$ decrease as powers of $1/\ln(E)$
and can be calculated using perturbation theory in $\alpha_s$.

To apply the OPE to ultracold atoms, the problem must be formulated 
in terms of a local quantum field theory.
Fermionic atoms with hyperfine spin states 
1, 2, and 3 that interact with large pair scattering lengths 
$a_{12}$, $a_{23}$, and $a_{13}$
can be described by a local quantum field theory 
with quantum fields $\psi_\sigma(\bm{r})$, $\sigma = 1, 2,3$.
The interaction term in the Hamiltonian density 
that gives the large scattering length $a_{12}$ is
\begin{equation}
{\cal H}_{\rm int} =
(\lambda_{0,12}/m) \psi_1^\dagger \psi_2^\dagger \psi_2 \psi_1.
\label{Hint}
\end{equation}
The bare coupling constant $\lambda_{0,12}$
depends on the ultraviolet momentum cutoff $\Lambda$:
$\lambda_{0,12} = 4 \pi/(1/a_{12}  - 2 \Lambda/\pi)$.
Similar interaction terms give the large scattering lengths
$a_{13}$ and $a_{23}$.
In Ref.~\cite{Braaten:2008uh}, the contact $C_{12}$ 
associated with atoms 1 and 2
was identified as the integral over space of the expectation value 
of a local operator:
\begin{equation}
C_{12} = \int \!\! d^3R 
\langle \lambda_{0,12}^2 \psi_1^\dagger \psi_2^\dagger 
          \psi_2 \psi_1(\bm{R}) \rangle  .
\label{contact-ZRM}
\end{equation}
Note that the contact density operator has an additional factor 
of the bare coupling constant compared to the interaction term in
Eq.~(\ref{Hint}).
In Ref.~\cite{Braaten:2008uh}, two of Tan's universal relations 
were rederived using the 
short-distance OPE.
The OPE
for $\psi_\sigma^\dagger$ and $\psi_\sigma$
was used to show that the momentum distributions 
for atoms of types 1 and 2 have identical 
$1/k^4$ tails whose coefficient is the contact $C_{12}$.
The OPE
for $\psi_1^\dagger \psi_1$ and $\psi_2^\dagger \psi_2$
was used to show that the contact density is proportional to 
the number of pairs of atoms of types 1 and 2 per (volume)$^{4/3}$.

The universal relations for rf spectroscopy in Eqs.~(\ref{sumrule-1}) 
and (\ref{tail-omega}) can be rederived and extended 
by using the short-time OPE.
We consider a many-body 
system consisting of ultracold atoms of types 1 and 2 only.
As far as the atoms are concerned,
rf spectroscopy with angular frequency $\omega$
proceeds through the action of the local operator
$\psi_3^\dagger \psi_2(\bm{r},t)$ on the many-body state.
The operator transforms an atom 
of type 2 into an atom of type 3 with energy larger by $\omega$.
The inclusive rate $\Gamma(\omega)$ for producing final states 
containing an atom of type 3 can be expressed formally 
in terms of the expectation value in the many-body state
of a product of operators:
\begin{eqnarray}
\Gamma(\omega) &=& \Omega^2 ~{\rm Im} \, i \!\!
\int \!\! dt~e^{i (\omega + i \epsilon) t} \int \!\! d^3R \int \!\! d^3r
\nonumber\\
&& \times 
\langle {\rm T} \psi_2^\dagger \psi_3(\bm{R} + \mbox{$\frac12$} \bm{r},t)~ 
 \psi_3^\dagger \psi_2(\bm{R} - \mbox{$\frac12$} \bm{r},0) \rangle.
\label{I-psipsi}
\end{eqnarray}

The Wilson coefficients in the OPE are determined by few-body 
physics and can be calculated diagrammatically
using methods described in Ref.~\cite{Braaten:2008bi}.
The lowest dimension operators that have nonzero 
expectation values in a state that 
contains no atoms of type 3 are $\psi_2^\dagger \psi_2$,
which has scaling dimension 3, 
and the contact density operator 
$\lambda_{0,12}^2 \psi_1^\dagger \psi_2^\dagger \psi_2 \psi_1$
which has scaling dimension 4.
The scaling dimensions of these and other operators can be deduced 
from the operator-state correspondence of conformal field theory
\cite{Nishida:2007pj}.
The Fourier transforms of the 
corresponding terms in the OPE are
\begin{eqnarray}
&& \int \!\! dt~e^{i \omega t} \int \!\! d^3r~
\psi_2^\dagger \psi_3(\bm{R} + \mbox{$\frac12$} \bm{r},t)~ 
 \psi_3^\dagger \psi_2(\bm{R} - \mbox{$\frac12$} \bm{r},0)
\nonumber \\
&&   \hspace{0.5cm} = 
(i/\omega)~\psi_2^\dagger \psi_2(\bm{R})
+ i W_{12}(\omega)~ 
\lambda_{0,12}^2 \psi_1^\dagger \psi_2^\dagger \psi_2 \psi_1(\bm{R})
\nonumber \\
&&   \hspace{1cm} 
+ \ldots.
\label{psipsi-OPE}
\end{eqnarray}
The Wilson coefficients
can be determined by matching the matrix elements 
of both sides between asymptotic incoming and outgoing 
few-atom states.
For $\psi_2^\dagger \psi_2$, one can use
single-atom states of type 2.
For the contact density operator, one can use 
two-atom scattering states consisting of atoms of types 1 and 2.
The matrix elements for these states can be
calculated using the analytic solution to the 2-body problem.  
The matrix element of the contact density operator 
is proportional to $1/|1/a_{12} + i k|^2$, where $k^2/m$
is the total energy of the scattering atoms. 
Its Wilson coefficient is
\begin{equation}
W_{12}(\omega) =  
\frac{a_{12}^{-1} - a_{13}^{-1}}{4 \pi m \omega^2}~
\frac{a_{12}^{-1} - \sqrt{-m \omega}} 
    {a_{13}^{-1} - \sqrt{-m \omega}}.
\label{C12}
\end{equation}
The OPE in Eq.~(\ref{psipsi-OPE}) is an asymptotic expansion for
large $|\omega|$ along any ray in the 
complex $\omega$ plane except possibly along the real $\omega$ axis, 
where the Wilson coefficients have poles and branch points.

The scaling of the Wilson coefficients at large $\omega$ is
determined simply by the scaling dimensions of the operators
and by symmetries. The rf transition operators on the left side 
of the OPE in Eq.~(\ref{psipsi-OPE}) each have scaling dimension 3.
Since the integral over nonrelativistic space and time 
has dimension $-5$, the total scaling dimension is 1.
On the right side of the OPE, the Wilson coefficient of 
an operator of dimension $d$ must have terms that scale
as $\omega^{(1-d-n)/2}(a^{-1})^n$, where $a$ is a scattering length.
Unless factors of $a^{-1}$ are required by a symmetry, 
the leading behavior is $\omega^{(1-d)/2}$. 
For $\psi_2^\dagger \psi_2$, which has scaling dimension 3, 
the leading behavior is correctly predicted to be $\omega^{-1}$.
The contact density operator has scaling dimension 4, 
so the leading behavior might be expected to be $\omega^{-3/2}$.
However the rf transition operators are invariant under a subgroup 
of an $SU(2)$ symmetry that is respected by the Hamiltonian if 
$a_{12}= a_{13}$ \cite{YB05}.  The contact density operator does not 
respect this symmetry, so $W_{12}(\omega)$ must have a factor of
$a_{13}^{-1} - a_{12}^{-1}$ that vanishes when $a_{12}= a_{13}$.
Thus its leading behavior is $\omega^{-2}$, in agreement with 
Eq.~(\ref{C12}).

The short-time OPE can be used to derive sum rules
for integrals over the rf frequency of the form
$\int_{-\infty}^\infty \!\! d \omega~F(\omega)\Gamma(\omega)$,
where $F(\omega)$ is a weight function that is analytic 
on the real $\omega$ axis. 
Upon inserting the expression in Eq.~(\ref{I-psipsi}) 
for $\Gamma(\omega)$ as a discontinuity in $\omega$ into the integral, 
it can be expressed as a line integral
over a contour that wraps around the real axis.
The sum rules in Eq.~(\ref{sumrules}) can be derived
by deforming the contour into a circle at infinity in the 
complex $\omega$ plane.  
The short-time OPE converges everywhere along the contour 
except possibly near the real axis.  
Inserting the OPE in Eq.~(\ref{psipsi-OPE}) into the expression for 
$\Gamma(\omega)$ in Eq.~(\ref{I-psipsi}) 
and then evaluating the contour integrals, 
we obtain the sum rules in Eq.~(\ref{sumrules}).

The short-time OPE can also be used to derive
the large-frequency behavior of the rf intensity.
Inserting the  OPE in Eq.~(\ref{psipsi-OPE}) into the 
expression for $\Gamma(\omega)$ in Eq.~(\ref{I-psipsi}), we obtain
\begin{equation}
\Gamma(\omega) \longrightarrow
\frac{\Omega^2 (a_{12}^{-1} - a_{13}^{-1})^2}
    {4 \pi \sqrt{m} \omega^{3/2} (a_{13}^{-2} + m \omega)}
~C_{12}.
\label{tail-omega123}
\end{equation}
It was pointed out in Ref.~\cite{PPS07} that $\Gamma(\omega)$
should decrease asymptotically like $\omega^{-5/2}$.
The coefficient of the contact in the $\omega^{-5/2}$ tail is a new result.
The analytic result for the rf transition rate $\Gamma(\omega)$ 
for the weakly-bound dimer associated with large positive $a_{12}$ \cite{CJ04}
agrees with Eq.~(\ref{tail-omega123}) in the limit 
$\omega \gg 1/(m a_{12}^2)$ if we use the fact that the contact 
of the dimer is $C_{12} = 8 \pi/a_{12}$ \cite{Tan}.
The result in Eq.~(\ref{tail-omega123}) for the tail in the 
rf transition rate relies on the scattering length $a_{13}$
being large compared to the range $r_0$.   If $a_{13}$ is not large, 
the corresponding result can be obtained by taking the limit 
$a_{13} \to 0$ in Eq.~(\ref{tail-omega123}),  
which gives Eq.~(\ref{tail-omega}).

Sum rules that are less sensitive to range effects can be obtained by 
using a weight function $F(\omega)$ that decreases as $\omega \to \infty$.
Using a Lorentzian centered at 
$\omega_0$ with half-width at half-maximum $\gamma$
gives a 2-parameter family of sum rules. 
Upon deforming the contour into a circle at infinity along which the 
integral vanishes, we are left with the contributions from the poles 
in $\omega$ at $\omega_0 \pm i \gamma$.
The convergence of the expansion for the sum rule is
governed by the convergence of the OPE at this complex frequency 
and is therefore insensitive to the breakdown of the OPE 
near the real axis.
In the limit $a_{13} \to 0$, this sum rule is
\begin{eqnarray}
&& \int_{-\infty}^\infty \!\!\! d\omega 
\frac{\gamma/\pi}{(\omega - \omega_0)^2 + \gamma^2}
\Gamma(\omega) 
= \frac{\Omega^2 \gamma}{\omega_\gamma^2} N_2
\nonumber \\
&&  \hspace{0.5cm}
+ \frac{\Omega^2 
       \big[(\omega_0^2 - \gamma^2) b_+
            + 2 \omega_0 \gamma b_-
            - 2 \omega_0 \gamma a_{12}^{-1} \big]}
      {4 \pi m \omega_\gamma^4}  C_{12}
\nonumber \\
&&  \hspace{0.5cm}
+ \ldots.
\label{PQWsumrule}
\end{eqnarray}
where $\omega_\gamma = (\omega_0^2 + \gamma^2)^{1/2}$
and $b_\pm = [m (\omega_\gamma \pm \omega_0)/2]^{1/2}$.
The analogous sum rule for $e^+ e^-$ annihilation into hadrons 
was derived in Ref.~\cite{Poggio:1975af}.
Sum rules of the form
$\int_0^{\omega_0} \!\! d \omega~F(\omega)\Gamma(\omega)$,
where $F(\omega)$ is a polynomial, are also insensitive 
to range effects if $\omega_0 \ll 1/(m r_0^2)$. 
The convergence of the expansion for the sum rule is governed 
by the convergence of the OPE on the circle $|\omega| = \omega_0$.
Sensitivity to the breakdown of the OPE 
near the real axis can be decreased by including a factor of 
$\omega - \omega_0$ in $F(\omega)$.

In the OPE in Eq.~(\ref{psipsi-OPE}), there are additional terms 
that have nonzero matrix elements in states that contain atoms of 
type 3.  The lowest dimension operator is the 23 analog 
of the 12 contact density operator in Eq.~(\ref{contact-ZRM}).
If the rf operators in Eq.~(\ref{psipsi-OPE}) are replaced by their 
hermitian conjugates, the local operators include 
$\psi_3^\dagger \psi_3$ and the 13 contact density operator.
There is also a 123 contact density operator proportional to 
$\psi_1^\dagger \psi_2^\dagger \psi_3^\dagger \psi_3 \psi_2 \psi_1$,
whose scaling behavior is governed by Efimov physics \cite{Braaten:2004rn}.
It has a complex scaling dimension whose real part is 5
and whose imaginary part is $\pm 2 s_0$, where $s_0 \approx 1.00624$.
Thus its Wilson coefficient will decrease asymptotically
like $\omega^{-2}$ multiplied by a log-periodic function of $\omega$
with a discrete scaling factor of approximately 515. 
The calculation of its Wilson coefficient requires the numerical 
solution to a 3-body problem.

The short-time OPE can be applied to other operators to derive sum rules 
and high-frequency tails for the associated spectral functions.
The OPE for $\psi_\sigma^\dagger$ and $\psi_\sigma$ provides universal
information about the spectrum of single-particle excitations.  
The OPE for density operators $\psi_\sigma^\dagger \psi_\sigma$ 
provides universal information about the spectrum of density fluctuations.
The OPE for two stress tensors provides constraints on the 
spectral functions that determine transport coefficients,
such as the viscosity. 

Universal relations involving the contact provide nontrivial examples
of aspects of many-body physics that are controlled by few-body
physics.  The OPE reveals these aspects by expressing observables in
terms of Wilson coefficients that are determined by few-body physics.
Of the known universal relations, most have been derived from the
analytic solution to the 2-body problem, but one has been derived from
the solution to the 3-body problem \cite{Werner:2010}.  The derivation
of new universal relations from numerical solutions to the 3-body and
higher few-body problems presents an interesting challenge.

{\it Note added:} The contact has recently been measured in
experiments with ultracold $^6$Li atoms \cite{Hu:2010} and 
$^{40}$K atoms \cite{Stewart:2010}. Several of the universal 
relations involving the contact have been verified experimentally.

\begin{acknowledgments}
This research was supported in part by the 
Army Research Office and the Air Force Office for Scientific Research 
and by the Department of Energy.
\end{acknowledgments}


%

\end{document}